\documentclass   [a4paper,epsfig,12pt]{article}
\usepackage{graphicx}
\begin{document}

\begin{center}

  {\bf  Semiclassical interpretation of microscopic processes}
 \footnote{Presented during the round-table discussion to the
conference "On the Present Status of Quantum Mechanics", held at
Mali Losinj (Croatia) in Sept. 6-9, 2005. }
             \\ [1mm]
    Milo\v{s} V. Lokaj\'{\i}\v{c}ek \\
 Institute of Physics, Academy of Sciences of the Czech Rep., 18221 Prague, Czech Republic
\end{center}
\vspace{5mm}


{\bf Abstract}

There are three upper limits ($2, 2\sqrt{2}, 2\sqrt{3}$) of the
Bell operator corresponding to different physical concepts:
classical, hidden-variable and quantum-mechanical. Only the
classical concept corresponding to the lowest limit has been
excluded by experimental data, while the other two should be
regarded as acceptable for the interpretation of EPR experiments
and all microscopic processes. A corresponding hidden-variable or
semiclassical model (based on the extended Hilbert space) will be
proposed and shortly described.
  \\


\vspace{12mm}

{\bf 1. Introduction}

The interpretation of EPR experiments \cite{ein} represents one of
the key problems of contemporary physics. It seemed that the
question would be fully answered on the basis of experimental data
when J. Bell \cite{bell} derived his famous inequalities. However,
the basic questions remained practically open, even if Aspect et.
al. \cite{asp} showed that the given inequalities have been
violated in the experiment and the data have been practically in
agreement with quantum-mechanical predictions.

We should like to show that the given situation has been burdened
from the very beginning by two following facts:

- It has been believed that Bell's inequalities have been derived
without any important assumption; however,  a far-reaching
assumption has been involved (see Sec. 2).

- It was stated by Belinfante \cite{belif} that the prediction of
any hidden-variable theory for photon transmission through a
polarizer pair should differ strongly from the quantum-mechanical
one, which has been based on an assumption contradicting the
reality (see Sec. 3).

We will show in Sec. 2 that there are different limits of
expectation values of Bell operator that correspond to divers
physical concepts. Only the classical limit (being regarded
mistakenly until now as the hidden-variable one) is clearly
excluded by experimental data, while the other two
(hidden-variable and quantum- mechanical) must be denoted as
acceptable. In Sec. 3 the correct hidden-variable formula for
photon transmission through two polarizers will be derived. And
finally, a corresponding mathematical hidden-variable model will
be proposed and described in Sec. 4.
\\ [3mm]

{\bf 2. Different limits of Bell operator}

Bell (and also his followers) has derived the given inequalities
for a combination of different coincidence probabilities for two
spin particles (photons), having included an assumption concerning
the individual transition probabilities. We will show now its
consequence specifying various conditions that lead to divers
limit values and correspond to different physical concepts. It
will be done in the language of the so called Bell operator
obtained by substituting individual probabilities by basic
operators that correspond to individual measurement acts (see,
e.g., \cite{hill}).

The Bell operator $B$ is defined in the Hilbert space
 \begin{equation}
     {\mathcal H}\;=\; {\mathcal H}_a \otimes {\mathcal H}_b
                         \label{tens}
 \end{equation}
where the subspaces ${\mathcal H}_a$ and ${\mathcal H}_b$
represent individual measuring devices (polarizers) in the
coincidence arrangement. It is then possible to write
 \begin{equation}
    B\;=\;a_1b_1+a_1b_2+a_2b_1-a_2b_2
 \end{equation}
where $a_j$ and $b_k$ are operators acting in subspaces ${\mathcal
H}_a$ and ${\mathcal H}_b$ and corresponding to measurements in
individual polarizers. It holds for the expectation values of
these operators
   $$ 0\;\leq\; |\langle a_j\rangle|, \;|\langle b_k\rangle|\; \leq \;1\, .  $$
The expectation values $|\langle B\rangle|$ of the Bell operator
may then possess  different upper limits according to the mutual
commutation relations of the operators $a_j$ and $b_k$.

 If it holds
   $$[a_1,a_2]\neq 0,\;[b_1,b_2]\neq 0\;,\;\;\mathrm {and\; also} \;\; [a_j,b_k]\neq 0\,,    $$
one can obtain by a rough estimate \cite{tsil}
   $$ \langle BB^+\rangle \leq 16 \;\; \mathrm {or}  \;\;\langle B\rangle \leq 4 \;. $$
However, after more detailed calculation one obtains \cite{revz}
 \begin{equation}
 \langle BB^+\rangle \leq 12, \;\;\;\;|\langle B\rangle| \leq 2{\sqrt 3}\; . \label{one}
 \end{equation}
If
 $$ [a_1,a_2]\neq 0,\;[b_1,b_2]\neq 0\;,\;\;\mathrm {but}\;\;\;[a_j,b_k]= 0\,,     $$
it holds
 \begin{equation}
   \langle BB^+\rangle \leq 8,
     \;\;\;\;|\langle B\rangle| \leq 2{\sqrt 2}  \;.  \label{two}
 \end{equation}
And finally, if all operators $a_j$ and $b_k$ commute mutually one
obtains
 \begin{equation}
   \langle BB^+\rangle \leq 4,
              \;\;\;\;|\langle B\rangle| \leq 2 \; ;  \label{three}
 \end{equation}
the same limit being obtained also if all operators at least in
one of the subspaces ${\mathcal H}_a$ and ${\mathcal H}_b$ commute
mutually \cite{revz}.

These three different limits correspond to divers physical
alternatives:

(i) In contradiction to common opinion the last limit
(\ref{three}) corresponds to the conditions of classical physics.
The assumption (interchange of transmission probabilities between
different pairs of photons) introduced by Bell \cite{bell} (see
also, e.g., \cite{clau}) has been equivalent in its consequences
to that used by von Neumann \cite{neu} in 1932, which has been
mentioned recently also by Malley \cite{mall}.

(ii) The limit (\ref{two}) represents the properties of an
hidden-variable alternative; it is not the limit (\ref{three}) as
commonly assumed until now.  The corresponding hidden parameters
involve the variables characterizing the properties of measuring
devices (such as exact coordinates of individual atoms and
similarly), the values of which may be hardly exactly specified,
even if they influence measurement results.

(iii) As to the limit (\ref{one}) it represents the case when the
results of both the measuring devices are being mutually
influenced; i.e., the case of the orthodox quantum mechanics. In
the past it has been stated that only the quantum-mechanical
alternative has been allowed by experimental data.

However, it is only the classical limit that has been excluded by
experimental data. As to the hidden-variable alternative it does
not contradict the experiment and  may be brought to agreement
(see Sec. 3) with experimentally established coincidence
polarization data (obtained, e.g., by Aspect et al. \cite{asp}).
\\

{\bf  3. Photon transmission through a polarizer pair and
hidden-variable prediction}

The last statement seems, however, to be in disagreement with that
of Belinfante (see \cite{belif}, p. 284) that the hidden-variable
prediction must differ significantly from the experimental data
obtained for two polarizers and represented in principle by Malus
law
 $$M(\alpha) \cong \cos^2\alpha$$
where $\alpha$ is the mutual deviation of polarizer axes. However,
Belinfante started from a non-physical assumption. In fact,
already a very simple hidden-variable alternative may give
approximately required results, as will be shown in the following.

The transmission probability of unpolarized beam through a
polarizer pair in the hidden-variable description may be expressed
by
\begin{equation}
 P(\alpha) \;=\;\int^{\pi/2}_{-\pi/2}d\lambda\;p_1(\lambda)\;p_1(\alpha-\lambda)
                         \label{palf}
\end{equation}
where $\lambda$ are the spin (polarization) deviations  of
individual incoming photons from the axis of the first polarizer
and $p_1(\lambda)$ is the corresponding transition probability
through the polarizer; parameter $\lambda$ being introduced by
Bell. Formula (\ref{palf}) is valid for both one-side and
coincidence arrangements of two polarizers.

We can then easily obtain
        $$ P(\alpha)/P(0) \;\cong\; M(\alpha)  $$
if, e.g.,
     \[ p_1(\lambda) \;=\; 1- \frac{1-\exp(-(a|\lambda|)^{e})}{(1 +
   c\exp(-(a|\lambda|)^{e})},  \;\;\;\;a=1.95, \;\;e=3.56, \;\;c=500\; ;   \]
$(a\lambda), e, c\;$ being dimensionless numbers. Function
$p_1(\lambda) \;(=p_1(-\lambda))$ is represented by full line and
$P(\alpha)$ by dashed line in Fig. 1; $\lambda$ or $\alpha$ being
shown on abscissa. It holds also for the intensity transmitted
through the first polarizer
 \[  I_1/I_0 \;=\; \frac{1}{\pi}\int^{\pi/2}_{-\pi/2}d\lambda\;p_1(\lambda)
                               \;\cong\; 0.45 .  \]
 Belinfante came to his conclusion when he put
quite arbitrarily: $p_1(\lambda) \;=\;\cos^2\lambda$; more
detailed explanation of the given problem being found in
\cite{los}.

Thus, both the previous arguments against the hidden-variable
theory have been removed and nothing seems to prevent the
polarization EPR experiments from being described with the help of
the given theoretical alternative. And the last question should be
put: Is it possible to propose a corresponding mathematical model
that would fulfill all required properties? Such a model has been
already proposed \cite{kunlo,lokun}; the goal having been reached
by adapting the basic assumptions of the standard
quantum-mechanical model, which will be mentioned shortly in the
next section.
\\

{\bf  4. Hidden-variable (semiclassical) model of EPR experiments}

It is possible to say that the standard quantum-mechanical
mathematical model is based on the following three assumptions:

 (i) The description of a physical system is given by the complex function
$\Psi({\tilde x},t)$ obtained as a solution of the time-dependent
Schr\"{o}dinger function
\begin{equation}
   i \frac{\partial}{\partial t}\Psi({\tilde x},t)
\;=\; H\:\Psi({\tilde x},t)      \label{schr}
\end{equation}
 where the Hamiltonian
\begin{equation}
 H \;=\; \sum^N_{k=1}\sum^3_{j=1}\frac{p_{j,k}^2}{2m_k} +
                   V(\tilde x)     \label{hami}
\end{equation}
 represents its total energy;

(ii)  Individual states $\Psi({\tilde x},t)$ are represented by
vectors in the Hilbert space ${\mathcal H}$ spanned on the
eigenfunctions of the Hamiltonian:
         \[ H\, u_E({\tilde x}) \;=\; E\, u_E({\tilde x}) \;.  \]
The expectation physical values are then established with the help
of standard rules.

(iii) A physical meaning is attributed to the mathematical
superposition principle holding in such a space. Only in some more
complex physical systems the so-called superselection rules have
been applied to, even if no theoretical reason for different
handling has been given in principle.

It is possible to state that the measurement postulate proposed by
von Neumann and involving the so called wave collapse has been the
direct consequence of these three mathematical conclusions.

A corresponding mathematical hidden-variable model has been then
obtained when the first assumption has conserved, the last
assumption has been abandoned (superselection rules being extended
inside ${\mathcal H}$), and the second assumption substituted by:

(ii) The Hilbert space has been extended (twice doubled in the
general case) as proposed and described to a greater detail in
Refs. \cite{kunlo, lokun}; see also \cite{lok04}. There is also
one-to-one correspondence between a vector in the Hilbert space
and an actual experimental state in the framework of such a model.

However, as to the EPR problem (and all microscopic collision
processes) a much simpler Hilbert space structure is sufficient
that corresponds to the scattering theory proposed by Lax and
Phillips \cite{lax,lax2} many years ago. The Hilbert subspace
$\mathcal {H}$ consists in such a case of two subspaces
 $$ {\mathcal{H}} = {\mathcal{D}_-} \oplus {\mathcal{D}_+}  $$
where each subspace $\mathcal{D}_-$ and $\mathcal{D}_+$ is spanned
on one set of Hamiltonian eigenfunctions; both the subspaces being
related mutually with the help of evolution operator
$U(t)=\exp(-iHt)$. The evolution goes from incoming states
($\mathcal{D}_-$) to outgoing states ($\mathcal{D}_+$) in an
irreversible way; it holds, e.g.,
 $\langle U(t)\mathcal{D}_+|\mathcal{D}_-\rangle\,=\,0$.

In the description of coincidence EPR system the time evolution in
individual subspaces may be, of course, neglected. The incoming
and outgoing states may be represented by simple vectors in
agreement with Eq. (\ref{tens}). Only the transition of an
incoming state to an outgoing one is important, which may be
described by
 the products of operators $\;a_j\,b_k\;$ representing the transition
probabilities of corresponding photon-pair states from
$\mathcal{D}_-$ to $\mathcal{D}_+$. The hidden-variable model of
polarization EPR experiments may be, therefore, brought to harmony
with the general mathematical structure proposed independently
already earlier \cite{kunlo,lokun}.
    \\

{\bf 5. Conclusion}

To conclude we should like to stress that the correlation of
divers limit values of the Bell operator to different physical
concepts and theoretical approaches has opened quite new views how
to interpret EPR experiments (and other microscopic phenomena) and
to contribute to looking for new ways in physical thinking. It is
evident that the hidden-variable (semiclassical) theory cannot be
excluded from further considerations concerning the EPR problem,
while the standard quantum-mechanical description seems to be
unnecessarily broad. A definite decision between these two
alternatives must be done, however, on the basis of other
experiments.

I should like to appreciate very much numerous and valuable
discussions with my colleagues V. Kundr\'{a}t and P. Kundr\'{a}t.
\\




\vspace{2cm}

\begin{figure}[htb]
\begin{center}
\includegraphics*[scale=.35, angle= -90]{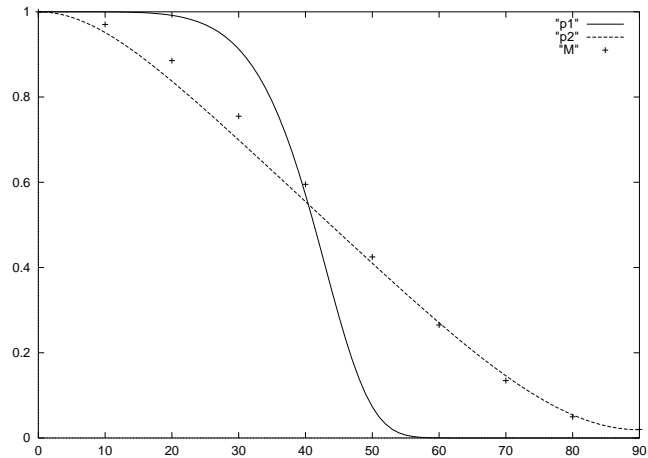}
\vspace{-2mm}
\caption   { {\it     Malus law and hidden variables;  $p_1(\lambda)$ - full line,
     $P_2(\alpha)$ - dashed line; Malus law  - individual points. }}
\end{center}
 \end{figure}

\end{document}